\magnification 1200
\font\tenmsb=msbm10   
\font\sevenmsb=msbm7
\font\fivemsb=msbm5
\newfam\msbfam
\textfont\msbfam=\tenmsb
\scriptfont\msbfam=\sevenmsb
\scriptscriptfont\msbfam=\fivemsb
\def\Bbb#1{\fam\msbfam\relax#1}
\let\nd\noindent 
\def\qed{\hbox{\hskip 6pt\vrule width6pt height7pt depth1pt \hskip1pt}}
\def\natural{{\rm I\kern-.18em N}}

\def\I{{\Bbb I}}
\def\J{{\Bbb J}}
\def\integer{{\rm Z\kern-.32em Z}}
\def\chix{{\raise.5ex\hbox{$\chi$}}}
\def\gammax{{\raise.5ex\hbox{$\gamma$}}}

\def\real{{\rm I\kern-.2em R}}

\def\Q{{\Bbb Q}}
\def\O{{\cal O}}
\def\complex{\kern.1em{\raise.47ex\hbox{
            $\scriptscriptstyle |$}}\kern-.40em{\rm C}}

\def\I{{\Bbb I}}
\def\vs#1 {\vskip#1truein}
\def\hs#1 {\hskip#1truein}

  \hsize=6truein        \hoffset=.25truein 
  \vsize=8.8truein      
  \pageno=1     \baselineskip=12pt
  \parskip=3 pt         \parindent=20pt
  \overfullrule=0pt     \lineskip=0pt   \lineskiplimit=0pt
  \hbadness=10000 \vbadness=10000 
\pageno=0

\footline{\ifnum\pageno=0\hss\else\hss\tenrm\folio\hss\fi}
\hbox{}
\vskip 1truein\centerline{{\bf ON ANGLES WHOSE SQUARED TRIGONOMETRIC}}
\vs.1
\centerline{{\bf FUNCTIONS ARE RATIONAL}}
\vskip .3truein\centerline{by}
\vskip .2truein
\centerline{{John H.\ Conway${}^1$ 
\footnote*{Research supported in part by NSF Grant No. DMS-9701444}}
\ \ {Charles Radin${}^2$}
\footnote{**}{Research supported in part by NSF Grant No. DMS-9531584 and
\hfill\break \indent Texas ARP Grant 003658-152\hfil}\ \ 
and\ \ Lorenzo Sadun${}^3$
\footnote{***}{Research supported in part by NSF Grant No. DMS-9626698 and
\hfill\break \indent Texas ARP Grant 003658-152\hfil}}
\vskip .2truein\centerline{\vbox{
${}^1$\ \ Department of Mathematics, Princeton University, Princeton, NJ\ \ 08544
\vskip.1truein
${}^{2,3}$\ \ Department of Mathematics, University of Texas, Austin, TX\ \ 78712}}
\vs.5
\centerline{{\bf Abstract}}
\vs.1 \nd
We consider the rational linear relations between real numbers whose
squared trigonometric functions have rational values, angles we call
``geodetic''. We construct a convenient basis for the vector space
over $\Q$ generated by these angles. Geodetic angles and rational
linear combinations of geodetic angles appear naturally in
Euclidean geometry; for illustration we apply our results to
equidecomposability of polyhedra.
\vfill\eject 

\nd
{\bf 0. Introduction}

Many well known geometric objects involve angles that are irrational
when measured in degrees or are irrational multiples of $\pi$ in
radian measure.  For instance we might mention the dihedral angle
$\alpha\ (\approx 70^\circ 31' 44'')$ of the regular tetrahedron,
whose supplement $(\approx 109^\circ 28' 16'')$ is known to chemists
as the carbon valence bond angle. A goodly number of these angles have
the property that their six trigonometric functions have rational
squares. For instance,
$$ 
\sin^2\alpha={8\over 9},\ \ \cos^2\alpha={1\over 9},\ \
\tan^2\alpha=8,\ \ \cot^2\alpha={1\over8},\ \
\sec^2\alpha=9,\ \ \csc^2\alpha={9\over 8} \eqno 1)
$$
and, for the dihedral angle $\beta\ (\approx 144^\circ 44' 8'')$ of the
cuboctahedron,
$$ \sin^2\beta={2\over 3},\ \ \cos^2\beta={1\over 3},\ \
\tan^2\beta={2},\ \ \cot^2\beta={1\over 2},\ \
\sec^2\beta={3},\ \ \csc^2\beta={3\over 2}. \eqno 2)
$$
There are many additive relations between angles of this kind; for
instance $\alpha$ and $\beta$ satisfy $\alpha + 2 \beta = 2 \pi$. In this
paper we essentially classify all such additive relations. 

To be precise, we shall say that $\theta$ is a ``pure geodetic angle''
if any one (and therefore each) of its six squared trigonometric
functions is rational (or infinite), and use the term ``mixed geodetic
angle'' to mean a linear combination of pure geodetic angles with
rational coefficients. The mixed geodetic angles form a vector space
over the rationals and we shall find an explicit basis for this space.
Finding a basis is tantamount to classifying all rational linear
relations among mixed geodetic angles.  By clearing denominators,
rational linear relations among mixed geodetic angles are easily
converted to additive relations among pure geodetic angles.

Another aim of this paper is to introduce an elegant notation for
these angles which we hope will find general acceptance. Namely, if
$0\le r\le 1$ is rational we define

$$\angle r=\sin^{-1}\sqrt{r}. \eqno 3)$$

\nd (We feel free to write these angles in either degrees or radians.) The
well-known particular cases are:

$$\angle 0=0^\circ =0,\ \angle {1\over 4}=30^\circ={\pi\over 6},\
\angle {1\over 2}=45^\circ = {\pi\over 4},\ 
\angle {3\over 4}=60^\circ={\pi\over 3},\
\angle 1=90^\circ={\pi\over 2}. \eqno 4)
$$
We extend this notation for all integers $n$, by writing
$$
\angle n= 90n^\circ={n\pi \over 2},\ \ \angle (n+r)=\angle n+\angle
r. \eqno 5)$$

Our basis contains certain angles $\langle p\rangle_d$ for prime $p$
and square-free positive $d$. If $p>2$ or if $p=2$ and $d\equiv 7$ (mod 8),
then $\langle
p\rangle_d$ is defined just when $-d$ is congruent to a square modulo
$p$ and is found as follows. Express $4p^s$ as $a^2+db^2$ for the
smallest possible positive $s$. Then

$$\langle p\rangle_d={1 \over s}\angle {db^2\over 4p^s}=
{1\over s}\sin^{-1}\sqrt{{db^2\over 4p^s}}
={1\over s}\cos^{-1}\sqrt{{a^2\over 4p^s}}
={1\over s}\tan^{-1}\sqrt{{db^2\over a^2}}. \eqno 6)
$$
The expression is unique except when $d=1$ or 3, when we make it so
by demanding that $b$ be even (if $d=3$) or divisible by four (if $d=1$). 
(Some exercise in the notation is provided in
Tables 1 and 2, which show the first few elements in the basis.) 
Our main result is then

\nd {\bf Theorem 1.} Every pure geodetic angle is uniquely expressible as
a rational multiple of $\pi$ plus an integral linear combination of
the angles $\langle p\rangle_d$. So the angles $\langle p\rangle_d$,
supplemented by $\pi$ (or $\angle 1$ or $1^\circ$), form a basis for
the space of mixed geodetic angles.

It is easy to find the representation of any pure geodetic angle $\theta$
in terms of the basis. 

\nd {\bf Theorem 2.} If $\tan \theta={b\over a}\sqrt{d}$ for integers
$a,b,d$, with square-free positive $d$ and with relatively prime $a$ and $b$,
and if the prime factorization of $a^2+db^2$ is $p_1 p_2 \cdots p_n$ (including
multiplicity), then we have
$$
\theta=t\pi\pm \langle p_1\rangle_d \pm \langle p_2\rangle_d\pm \cdots
\pm \langle p_n \rangle_d \eqno 7)
$$
for some rational $t$. 

(We note that the denominator of $t$ will be a divisor of the class
number of $\Q(\sqrt{-d})$.)

For example, for $\tan \theta={5\over 4}\sqrt{3}$ we find
$a^2+db^2=16+75=91=7\cdot 13$ and indeed $\theta
=n \pi-\langle 7\rangle_3-\langle 13\rangle_3$.

Our results have an outstanding application. In 1900 Dehn [Deh] solved
Hilbert's 3$^{rd}$ problem by giving a necessary condition for the
mutual equidecomposability of polyhedra in terms of their dihedral
angles, from which it follows easily that there are tetrahedra of
equal volume which are not equidecomposable. In 1965 Sydler proved
[Syd] that Dehn's criterion is also sufficient.  For polyhedra with
geodetic dihedral angles our Theorem 1 makes the Dehn-Sydler criterion
effective. At the end of this paper we shall apply our theory to the
non-snub Archimedean polyhedra (whose dihedral angles are all
geodetic.)
\vs.2 \nd
{\bf 1. Angles with polyquadratic tangents and the Splitting
Theorem.}

The addition formula for tangents enables us to show that the tangent
of any sum of pure geodetic angles is a ``polyquadratic
number'', that is a number of the form $\sqrt{a} + \sqrt{b} + \sqrt{c}
+\cdots$, with $a,b,c\cdots$ rational. For instance, if
$\tan\alpha=\sqrt{2}/2$ and
$\tan\beta=\sqrt{3}/3$, then
$$\eqalign{\tan(\alpha+\beta)&={\sqrt{2}/2+\sqrt{3}/3 \over 
1-\sqrt{2}\sqrt{3}/6}
={4\over 5}\sqrt{2} +{3\over 5}\sqrt{3}; \cr 
\tan(\alpha -\beta)&={4\over 5}\sqrt{2} -{3\over 5}\sqrt{3}.\cr} \eqno 8)$$

We now suppose the sum of a number of pure geodetic angles is an
integral multiple of $\pi$; let us say
$\alpha_1+\alpha_2+\cdots+\beta_1+\beta_2+\cdots =m\pi$, where we have
chosen the notation so that the tangents of $\alpha_1,\alpha_2,\cdots$
are in $\Q(\sqrt{d_1},\cdots, \sqrt{d_n})$ and those of
$\beta_1,\beta_2,\cdots$ are in $\sqrt{d}\Q(\sqrt{d_1},\cdots,
\sqrt{d_n})$, where $\sqrt{d}\notin\Q(\sqrt{d_1},\cdots,
\sqrt{d_n})$. Then, by the addition and substraction formulas for
tangents,
$\tan(\alpha_1+\alpha_2+\cdots+\beta_1+\beta_2+\cdots)$ and
$\tan(\alpha_1+\alpha_2+\cdots-\beta_1-\beta_2-\cdots)$ will be of the
form $a+b\sqrt{d}$ and $a-b\sqrt{d}$ where $a,b\in
\Q(\sqrt{d_1},\cdots, \sqrt{d_n})$.  But by assumption $a+b\sqrt{d}=0$,
so $a=b=0$, from which it follows that
$\alpha_1+\alpha_2+\cdots-\beta_1-\beta_2-\cdots$ is also an integral
multiple of $\pi$. Adding and subtracting we deduce that the two
subsums $\alpha_1+\alpha_2+\cdots$ and $\beta_1+\beta_2+\cdots$ are
integral multiples of $\pi/2$. Combining this argument with induction
on $n$
we obtain

\nd {\bf Theorem 3 (The Splitting Theorem).} If the value of a 
rational linear combination of pure geodetic angles is a rational
multiple of $\pi$ then so is the value of its restriction to those
angles whose tangents are rational multiples of any given square root.
\vs.1 
We remark that the same method can be used to show that any angle
whose tangent is polyquadratic is a mixed geodetic angle. For suppose
$\alpha$ is an angle whose tangent is polyquadratic, with
$\tan\alpha\in \sqrt{d_0} \Q(\sqrt{d_1},\ldots,\sqrt{d_n})$. So
$\tan\alpha=z_1+z_2\sqrt{d_n}$, where $z_j\in
\sqrt{d_0}\Q(\sqrt{d_1},\ldots,\sqrt{d_{n-1}})$. Choose $\alpha'$ such that
$\tan\alpha'=z_1-z_2\sqrt{d_n}$, and define $\gamma=\alpha+\alpha'$
and $\delta=\alpha-\alpha'$. It follows that

$$\eqalign{\tan\gamma=&
{\tan\alpha + \tan\alpha'\over 1-\tan\alpha \tan\alpha'}={2z_1\over
1-{z_1}^2+d_n{z_2}^2}\in \sqrt{d_0}\Q(\sqrt{d_1},\ldots,\sqrt{d_{n-1}}); \cr
\tan\delta=&{\tan\alpha - \tan\alpha'\over 1+\tan\alpha \tan\alpha'}={2z_2\sqrt{d_n}\over
1+{z_1}^2-d_n{z_2}^2}\in \sqrt{d_0d_n}\Q(\sqrt{d_1},\ldots,\sqrt{d_{n-1}}).} \eqno 9)
$$

Repetition of this technique justifies our claim. We leave to the
reader the exercise of applying this technique to the case of
$\tan\alpha=\sqrt{6}+\sqrt{3}+\sqrt{2}+{1}$, obtaining $4\alpha = 
\angle (1+ {441\over 457})+\angle {432\over 457}+
\angle {96\over 457}+\angle {2592\over 4113}$.
\vs.2 
\nd {\bf 2. St\o rmer theory and its generalization.}

The Splitting Theorem reduces the study of the rational linear relationships
between angles of the form $\tan^{-1}({b\over a}\sqrt{d})$ to those
with a fixed $d$. These angles are the arguments of algebraic integers
$a+b\sqrt{-d}$, and their theory is essentially the factorization
theory of numbers in $\O_d$, the ring of algebraic integers of 
$\Q(\sqrt{-d})$ [Pol]. The method was first used by C.\
St\o rmer [St\o] (in the case $d=1$) who classified the additive
relations between the arctangents of rational numbers using the unique
factorization of Gaussian integers. (See also [Con].) We recall St\o
rmer's analysis of the case $d=1$ and then generalize it to arbitrary $d$,
which will prove Theorems 1 and 2.

It is known that the Gaussian integers have unique factorization up to
multiplication by the 4 units: $1,\ -1,\ i,\ -i$.  It is also known
how each rational prime $p$ factorizes in the Gaussian
integers. Namely: 1) if $p\equiv -1$ (mod 4) then $p$ remains prime;
2) if $p\equiv +1$ (mod 4), then $p=a^2+b^2$ is the product of the distinct
Gaussian primes $a+ib$ and $a-ib$ (for uniqueness we choose $a$ odd,
$b$ even, both positive); and 3) $2=-i(1+i)^2$ ``ramifies'', that is
to say it is (a unit times) the square of a Gaussian prime.

Now let $\kappa=\pi_1\pi_2\pi_3\cdots$ be the prime factorization of a
Gaussian integer $\kappa$. Then plainly $\arg(\kappa)\equiv \arg(\pi_1)+
\arg(\pi_2)+\arg(\pi_3)+\cdots$ (mod $2\pi$). So the arguments of 
Gaussian primes (together with $\pi$) span the subspace of
mixed geodetic angles generated by the pure geodetic angles with rational
tangent. However,
\vs.1
1) If $p=4k-1$, then $\arg(p)=0$ and can be ignored.
\vs.05
2) If $p=4k+1=a^2+b^2$, then $\arg(a+bi)+\arg(a-bi)=0$. We define
\vs0 \nd \hs.53
$\langle p \rangle_1=\arg(a+bi)=-\arg(a-bi)$.
\vs.05
3) $\arg(1+i)=\pi/4$.
\vs.05
4) The arguments of the units are multiples of $\pi/2$, and so are
multiples of 
\vs0 \nd \hs.56
$\arg(1+i)$.
\vs.05 \nd
Thus the argument of any Gaussian integer is an integral linear
combination of $\pi/4$ and the angles $\langle p\rangle_1$, 
with $p\equiv 1$ (mod 4).

It is also easy to see that the numbers $\pi$ and the $\langle
p\rangle_1$'s are rationally independent. Otherwise some integral linear
combination of them would be an integral multiple of $\pi$. But
suppose for instance that $2\langle p_1\rangle_1-3\langle
p_2\rangle_1+5\langle p_3\rangle_1=0$.  The left hand side is the
argument of $\pi_1^2\bar\pi_2^3\pi_3^5$ which must therefore be a real
number and hence should be equal to its conjugate
$\bar\pi_1^2\pi_2^3\bar\pi_3^5$. But this contradicts the unique
factorization of Gaussian integers.

The analogue of the St\o rmer theory for the general case is complicated
by the fact that some elements of $\O_d$ may not have unique
factorization.  However the ideals do. Instead of assigning arguments to
numbers, we simply assign an angle to each ideal $\I$  by the rule
$$ \arg(\I) = \cases{\arg(\kappa) & if $\I = (\kappa)$ is principal; \cr
{1 \over s} \arg(\I^s) & if $\I$ is not principal,} \eqno 10)
$$
where $s$ is the smallest exponent for which $\I^s$ is a principal
ideal, and $(\kappa)$ denotes the principal ideal generated by $\kappa$.  
Recall that for every $d$ the ideal class group is finite, 
so such an $s$ exists for every ideal, and $s$ divides the class number
of $\O_d$.  Since the generator of a principal ideal is unique
up to multiplication by a unit, we take the argument of an ideal to
be defined only modulo the argument of a unit divided by the class number of 
$\O_d$. This ambiguity is always a rational multiple of $\pi$.

Let $c_d$ be the class number of $\O_d$.  For every ideal $\I$,
principal or not, we have that the ideal argument $\arg(\I)$ is equal
to $1/c_d$ times the (ordinary) argument of the generator of the
principal ideal $\I^{c_d}$, up to the ambiguity in the definition of
ideal arguments.  It follows that, for general ideals $\I$ and $\J$,
$$ \arg(\I\J) = \arg(\I) + \arg(\J), \eqno 11)
$$
(modulo the ambiguity) since the (ordinary) argument of the generator of
$(\I\J)^{c_d}$ is the argument of the generator of $\I^{c_d}$
plus the argument of the generator of $\J^{c_d}$ (mod $\pi$).
Thus the argument of any ideal (and in particular
the principal ideal generated by any algebraic integer) is an
integral linear combination of the arguments of the prime ideals (modulo
the ambiguity).  
What remains is to determine the nontrivial arguments of prime ideals.
As in the case $d=1$, there will be one such angle for each rational
prime $p$ for which $(p)$ splits as the product of distinct ideals.

We illustrate the procedure by working in
$\O_5$, for which the ideal factorizations of the first few
rational primes are:
$$ 
\eqalign{ (2)=& (2,1+\sqrt{-5})^2 \cr 
(3)=& (3,1+\sqrt{-5})(3,1-\sqrt{-5}) \cr 
(5)=& (\sqrt{-5})^2 \cr 
(7)=& (7,3+\sqrt{-5})(7,3-\sqrt{-5}) \cr 
(11)= & (11) \cr 
(13)= & (13) \cr
(17)= & (17) \cr 
(19)= & (19) \cr 
(23)= & (23,22+3\sqrt{-5})(23,22-3\sqrt{-5}) \cr 
(29)= & (3+2\sqrt{-5})(3-2\sqrt{-5}).} \eqno 12) 
$$ 
(This list may be obtained using Theorems 5 and 6, below.)
Here $(x,y)$ denotes the
ideal generated by $x$ and $y$.  The reader will see that (2) ramifies as
the square of a non-principal ideal and (5) as the square of a
principal ideal, (3), (7) and (23) split into products of
non-principal ideals, (29) splits as the product of distinct principal
ideals, while (11), (13), (17) and (19) remain prime.
Notice that, as the example shows, every ideal of $\O_d$ can be generated
by at most two numbers, and $(p)$ can be written as a product of at most
two prime ideals, for any rational prime $p$.

As in the St\o rmer case the principal ideals generated by 
rational primes that remain prime have
argument zero and can be ignored.  We also ignore those that ramify,
since their angles will be rational multiples of $\pi$. Otherwise we
define $\langle p\rangle_d$ to be the argument of one of the two ideal
factors of $(p)$, making it unique by requiring $0<\langle
p\rangle_d<\pi/2$ if the factors of $(p)$ are principal and $0<\langle
p\rangle_d<\pi/4$ if not.

To illustrate this we determine $\langle 3\rangle_5$. The ideal
factors of (3) are non-principal so we square them:
$$\eqalign{(3,1+\sqrt{-5})^2&=(3^2,3(1+\sqrt{-5}),(1+\sqrt{-5})^2)\cr
                       &=(9,3+3\sqrt{-5},-4+2\sqrt{-5}),} \eqno 13)
$$
which reduces to $(2-\sqrt{-5})$. 
Similarly, $(3,1-\sqrt{-5})^2=(2+\sqrt{-5})$. So
$\langle 3\rangle_5={1\over 2}\arg(2+\sqrt{-5})={1\over
2}\tan^{-1}({1\over 2}\sqrt{5})= {1\over 2}\angle {5\over 9}$.

In the general case ($d$ an arbitrary square-free positive integer) 
the previously described procedure
assigns an angle $\langle p\rangle_d$ to every rational prime $p$ for
which $(p)$ splits as the product of two distinct prime ideals $\I$
and $\J$.  Let $s$ be the smallest integer for which $\I^s$ (and therefore
$\J^s$) is
principal.  Recall that the elements of $\O_d$ are of the
form ${a\over 2}+{b\over 2}\sqrt{-d}$, where $a$ and $b$ are rational
integers.  If $d \not\equiv 3$ (mod 4), then $a$ and $b$ must be even; if
$d\equiv 3$ (mod 4) then $a$ and $b$ are either both even or both odd. We
can therefore write
$$
\I^s=\left ( {a\over 2}+{b\over 2}\sqrt{-d}\right ),\ \ 
\J^s=\left ( {a\over 2}-{b\over 2}\sqrt{-d}\right ), \eqno 14)
$$
where we can distinguish between $\I$ and $\J$ by supposing that
$a$ and $b$ are positive.
We take $\langle p\rangle_d={1\over s}
\tan^{-1}({b\over a}\sqrt{d})= {1 \over s} \angle {b^2d \over a^2 + b^2 d}$. 

The above defines the angles $\langle p\rangle_d$ uniquely for all $d$
other than 1 and 3, because then the only units are $\pm 1$, so that
the only generators of $\I^s$ and $\J^s$ are the four numbers $\pm
{a\over 2}\pm {b\over 2}\sqrt{d}$.  When $d=1$ we have the additional
unit $i$ which effectively allows us to interchange $a$ and $b$: we
then achieve uniqueness by demanding that the generators of $\I$ be
$a+bi$ with $a$ and $b$ positive integers with $b$ even. In the case
$d=3$ the field has six units and the corresponding condition is that
the generators of $\I$ should have the form $a+b\sqrt{-3}$ where $a$
and $b$ are positive integers.

\nd {\bf Theorem 4.} For fixed square-free positive $d$, consider 
the subspace of mixed geodetic angles generated by arctangents of
rational multiples of $\sqrt{d}$.  This subspace is spanned by
$\pi$ and the nonzero angles of the form $\langle p
\rangle_d$, where $p$ ranges over the rational primes for
which $(p)$ splits as a product of distinct ideals in $\O_d$.

\nd Proof. The proof is essentially that of the St\o rmer decomposition,
only substituting the arguments of ideals for the arguments of
algebraic integers.\qed

Combining Theorem 4 with the Splitting Theorem, and by the rational
independence of $\pi$ and the $\langle p \rangle_d$'s for any fixed $d$
(the independence can be proved similarly as it was shown in the case
$d=1$),
we obtain Theorem 1. 

\nd Proof of Theorem 2. Recall that if $\I$ is any ideal (principal
or not) in $\O_d$, then
$\I\bar \I$ is a principal ideal with a
positive integer generator that we call the norm of $\I$, and
that norms are multiplicative [Pol]. From this it follows that the prime ideals
are precisely the factors of $(p)$, with $p$ ranging over the rational 
primes, and that every prime ideal that is not generated by a rational 
prime has rational prime norm.

Now suppose that $\tan(\theta)={b\over a}\sqrt{d}$ with square-free positive $d$ 
and with relatively prime $a$ and $b$.  Consider the factorization of the 
principal ideal $\I=(a + b \sqrt{-d})$.  If $\I=\pi_1\pi_2\cdots\pi_n$,
where each ideal $\pi_i$ is prime, then none of the $\pi_i$'s are generated
by rational primes, insofar as $a$ and $b$ are relatively prime. Thus each
$\pi_i$ satisfies $\pi_i\bar \pi_i=(q_i)$ for some rational prime $q_i$. On
the other hand, we have
$$ (p_1)(p_2)\cdots(p_n) = (a^2 + b^2d) = 
\I \bar \I = \pi_1 \bar \pi_1 \cdots \pi_n \bar \pi_n.
\eqno 15)
$$
So, after a suitable permutation of the indices $1,\ldots,n$ on the right 
side of 15) we have $\pi_i
\bar \pi_i = (p_i)$, and so the argument of $\pi_i$ is 
$\pm \langle p_i \rangle_d$, for every $i$. But $\theta=\arg(a+b\sqrt{-d})=\arg(\I)$
is the sum of the arguments of the $\pi_i$'s, up to a rational multiple
of $\pi$ that comes from the ambiguity in the definition of the argument
of an ideal.\qed

All that remains is to identify the pairs
$(p,d)$ for which $\langle p
\rangle_d$ is defined. The following theorems
give the criteria. These criteria may be easily implemented, 
by hand for small $d$ 
and $p$, and by computer for larger values.  The theorems themselves
are standard results, and we leave the proofs to the reader.

\nd {\bf Theorem 5.} Let $p$ be an odd rational prime. The ideal $(p)$ 
of $\O_d$ splits as
a product of distinct ideals if and only if we can write
$$ 4 p^s = a^2 + b^2 d  \eqno 16) $$ 
for integers $a$ and $b$ (neither a multiple of $p$) 
and for an exponent $s$ that divides the
class number of $\O_d$.  If $d \not\equiv 3$ (mod 4), or if $d=3$, then the factor of
4 is unnecessary, and the criterion for splitting reduces to
$$ p^s = a^2 + b^2 d, \eqno 17) $$
for $a$ and $b$ nonzero (mod $p$).
The ideal $(p)$ is prime in 
$\O_d$ if and only if $-d$ is not equal
to a square modulo $p$.  If $(p)$ is not prime and does not split,
then $(p)$ ramifies. 

Theorem 5 gives criteria for all odd primes.  The prime $p=2$ is
somewhat different. Since both 0 and 1 are squares, every $-d$ is
congruent to a square modulo 2.  However, there are values of $d$ for
which $(2)$ is prime. 

\nd {\bf Theorem 6.}  If $d \not\equiv 3$ (mod 4), then $(2)$ ramifies
in $\O_d$.  If $d \equiv
3$ (mod 8), then $(2)$ is prime.  If $d\equiv 7$ (mod 8), then $(2)$
splits and we can write a power of 2 as $a^2 + b d^2$.
\vs.2 \nd
{\bf 3. Applications to the Dehn-Sydler criterion of Archimedean
polyhedra.}

The Dehn invariant of a polyhedron whose $i^{th}$ edge has length
$\ell_i$ and dihedral angle $\theta_i$ is the formal expression
$\sum_i\ell_iV[\theta_i]$ where the ``vectors'' $V[\theta_i]$ are
subject to the relations
$$V[r\theta+s\phi]=rV[\theta]+sV[\phi],\ \ V[r\pi]=0, \eqno 19)$$
for all rational numbers $r$ and $s$. The $V[\theta]$'s satisfy the same
rational linear relations satisfied by the angles $\theta$ in the
rational vector space they generate, together with the additional
relation $V[\pi]=0$; however we allow their coefficients to be
arbitrary real numbers.

If every dihedral angle $\theta$ of a polyhedron is geodetic we can 
write
$$\theta=r\pi+r_1\langle p_1\rangle_{d_1}+\cdots+r_j \langle
p_j\rangle_{d_j} \eqno 19)$$
for rational numbers $r,r_1,\cdots,r_j$, so
$$
V[\theta]=r_1V[\langle p_1\rangle_{d_1}]+\cdots+r_jV[\langle
p_j\rangle_{d_j}]. \eqno 20)
$$
If the edge lengths of the polyhedron are rational its Dehn invariant 
will then be a rational linear combination of the
$V[\langle p\rangle_{d}]$'s.

It can be easily checked that each face of an Archimedean polyhedron
(other than the snub cube and snub dodecahedron) is orthogonal to a
rotation axis of one of the Platonic solids, and the rotation groups
of all the Platonic solids are contained in the cube group ${\cal C}$
and icosahedral group ${\cal I}$.  It follows that the dihedral angles
of all these polyhedra are found among the supplements of the angles
between the rotation axes of ${\cal C}$ and ${\cal I}$.

We now concentrate on ${\cal I}$.
Let $\tau=(1+\sqrt{5})/2$ and $\sigma=\tau^{-1}=\tau-1$. 
The 12 vectors whose coordinates are cyclic permutations of $0,\pm
1,\pm\tau$ lie along the pentad axes. Similarly the 20 vectors obtained
by cyclicly permutating $\pm 1,\pm 1,\pm 1$ and $0,\pm\tau,\pm\sigma$ lie
along the triad axes, and the 30 cyclic permutations of $\pm2,0,0$ and
$\pm1,\pm\sigma,\pm\tau$ lie along the dyad ones. 
The cosines of the angles between the axes have the form $v\cdot
w/|v|\,|w|$ where $v$ and $w$ are chosen from these vectors. These
cosines are enumerated in Fig.$\,$1. The angles that correspond to
them are those shown in Fig.$\,$2, together with their supplements.

Table 3 gives the components of the Dehn invariants for the non-snub\hfill 
\break
Archimedean polyhedra of edge lengths 1. For instance the dihedral
angles of the truncated tetrahedron are $\pi - 2\langle 3\rangle_2$ at
six edges and $2\langle 3\rangle_2$ at the remaining 12, so that
its Dehn invariant is
$$
6V[\pi-2\langle 3\rangle_2]+12V[2\langle
3\rangle_2]=12V[\langle 3\rangle_2], \eqno 21)
$$
since $V[\pi]=0$. In the values we abbreviate $V[\langle p\rangle_d]$
to $\langle p\rangle_d$.

We note that the Dehn invariants of the icosahedron, dodecahedron and
icosidodecahedron with unit edge lengths, namely 
$60\langle 3\rangle_5, -30\langle 5\rangle_1$, and 
$30\langle 5\rangle_1-60\langle 3\rangle_5$, respectively,
have zero sum, so Sydler's theorem shows that it is possible to dissect
them into finitely many pieces that can be reassembled to form a large
cube. This might make an intriguing wooden puzzle if an explicit
dissection could be found. (We have no idea how to do this.)
\vs.2 \nd
{\bf 4. Angles with algebraic trigonometric functions.}

It is natural to consider a generalization of our theory that gives a
basis for the rational vector space generated by all the angles whose
six trigonometric functions are algebraic. What is missing here is the
analogue of our Splitting Theorem. If such an analogue were found,
the ideal theory would probably go through quite easily.

We ask a precise question: Does there exist an algorithm that finds
all the rational linear relations between a finite number of such
angles? The nicest answer would be one giving an explicit basis,
analogous to our $\langle p\rangle_d$.
\vfill \eject
\nd
{\bf References}
\vs.2
\nd
[Con]\ J.H.\ Conway, R.K.\ Guy, {\it The book of numbers}, Copernicus, New York, 1996.
\vs.1 \nd
[Deh]\ M.\ Dehn, Uber den Rauminhalt, 
{\it G\"ottingen Nachr.\ Math.\ Phys.} (1900), 345-354; {\it Math.\ Ann.}, 55 (1902), 465-478.
\vs.1 \nd
[Pol]\ H.\ Pollard and H.\ Diamond, {\it The theory of algebraic numbers}, Second edition, 
Carus Mathematical Monographs, 9,
Mathematical Association of America, Washington, D.C., 1975.
\vs.1 \nd
[Sah] C.-H.\ Sah, {\it Hilbert's third problem : scissors congruence},
Pitman, San Francisco, 1979.
\vs.1 \nd
[Syd]\ J.P.\ Sydler, Conditions n\'ecessaires et suffisantes pour l'\'equivalence
des poly\`edres l'espace euclidien \`a trois dimensions, {\it Comm.\ Math.\ Helv.}, 40
(1965), 43-80.
\vs.1 \nd
[St\o]\ C. St\o rmer: Sur l'application de la th\'eorie des nombres entiers complexes \`a
la solution en nombres rationnels $x_1,x_2\ldots x_n\ c_1c_2\ldots c_n, k$ de 
l'\'equation $c_1 \hbox{arc tg }x_1 + c_2 \hbox{arc tg }x_2 +\cdots c_n \hbox{arc tg }x_n = k{\pi\over 4}$, {\it Arch.\ Math.\ Naturvid.}, 19 No. 3 (1896).
\vfill \eject

%
%
\newdimen\FigSize	\FigSize=.9\hsize 
%
\newskip\abovefigskip	\newskip\belowfigskip
\gdef\epsfig#1;#2;{\par\vskip\abovefigskip\penalty -500
   {\everypar={}\epsfxsize=#1\noindent
    \centerline{\epsfbox{#2}}}%
    \vskip\belowfigskip}%
%
\newskip\figtitleskip
\gdef\tepsfig#1;#2;#3{\par\vskip\abovefigskip\penalty -500
   {\everypar={}\epsfxsize=#1\noindent
    \vbox
      {\centerline{\epsfbox{#2}}\vskip\figtitleskip
       \centerline{\figtitlefont#3}}}%
    \vskip\belowfigskip}%
%
\newcount\FigNr	\global\FigNr=0
\gdef\nepsfig#1;#2;#3{\global\advance\FigNr by 1
   \tepsfig#1;#2;{Figure\space\the\FigNr.\space#3}}%
%
%
%
\gdef\ipsfig#1;#2;{
   \midinsert{\everypar={}\epsfxsize=#1\noindent
	      \centerline{\epsfbox{#2}}}%
   \endinsert}%
%
\gdef\tipsfig#1;#2;#3{\midinsert
   {\everypar={}\epsfxsize=#1\noindent
    \vbox{\centerline{\epsfbox{#2}}%
          \vskip\figtitleskip
          \centerline{\figtitlefont#3}}}\endinsert}%
%
\gdef\nipsfig#1;#2;#3{\global\advance\FigNr by1%
  \tipsfig#1;#2;{Figure\space\the\FigNr.\space#3}}%
\newread\epsffilein    
\newif\ifepsffileok    
\newif\ifepsfbbfound   
\newif\ifepsfverbose   
\newdimen\epsfxsize    
\newdimen\epsfysize    
\newdimen\epsftsize    
\newdimen\epsfrsize    
\newdimen\epsftmp      
\newdimen\pspoints     
\pspoints=1bp          
\epsfxsize=0pt         
\epsfysize=0pt         
\def\epsfbox#1{\global\def\epsfllx{72}\global\def\epsflly{72}%
   \global\def\epsfurx{540}\global\def\epsfury{720}%
   \def\lbracket{[}\def\testit{#1}\ifx\testit\lbracket
   \let\next=\epsfgetlitbb\else\let\next=\epsfnormal\fi\next{#1}}%
\def\epsfgetlitbb#1#2 #3 #4 #5]#6{\epsfgrab #2 #3 #4 #5 .\\%
   \epsfsetgraph{#6}}%
\def\epsfnormal#1{\epsfgetbb{#1}\epsfsetgraph{#1}}%
\def\epsfgetbb#1{%
%
%
\openin\epsffilein=#1
\ifeof\epsffilein\errmessage{I couldn't open #1, will ignore it}\else
%
%
   {\epsffileoktrue \chardef\other=12
    \def\do##1{\catcode`##1=\other}\dospecials \catcode`\ =10
    \loop
       \read\epsffilein to \epsffileline
       \ifeof\epsffilein\epsffileokfalse\else
%
%
          \expandafter\epsfaux\epsffileline:. \\%
       \fi
   \ifepsffileok\repeat
   \ifepsfbbfound\else
    \ifepsfverbose\message{No bounding box comment in #1; using defaults}\fi\fi
   }\closein\epsffilein\fi}%
%
%
\def\epsfsetgraph#1{%
   \epsfrsize=\epsfury\pspoints
   \advance\epsfrsize by-\epsflly\pspoints
   \epsftsize=\epsfurx\pspoints
   \advance\epsftsize by-\epsfllx\pspoints
%
%
   \epsfxsize\epsfsize\epsftsize\epsfrsize
   \ifnum\epsfxsize=0 \ifnum\epsfysize=0
      \epsfxsize=\epsftsize \epsfysize=\epsfrsize
%
%
     \else\epsftmp=\epsftsize \divide\epsftmp\epsfrsize
       \epsfxsize=\epsfysize \multiply\epsfxsize\epsftmp
       \multiply\epsftmp\epsfrsize \advance\epsftsize-\epsftmp
       \epsftmp=\epsfysize
       \loop \advance\epsftsize\epsftsize \divide\epsftmp 2
       \ifnum\epsftmp>0
          \ifnum\epsftsize<\epsfrsize\else
             \advance\epsftsize-\epsfrsize \advance\epsfxsize\epsftmp \fi
       \repeat
     \fi
   \else\epsftmp=\epsfrsize \divide\epsftmp\epsftsize
     \epsfysize=\epsfxsize \multiply\epsfysize\epsftmp   
     \multiply\epsftmp\epsftsize \advance\epsfrsize-\epsftmp
     \epsftmp=\epsfxsize
     \loop \advance\epsfrsize\epsfrsize \divide\epsftmp 2
     \ifnum\epsftmp>0
        \ifnum\epsfrsize<\epsftsize\else
           \advance\epsfrsize-\epsftsize \advance\epsfysize\epsftmp \fi
     \repeat     
   \fi
%
%
   \ifepsfverbose\message{#1: width=\the\epsfxsize, height=\the\epsfysize}\fi
   \epsftmp=10\epsfxsize \divide\epsftmp\pspoints
   \vbox to\epsfysize{\vfil\hbox to\epsfxsize{%
      \includegraphics{#1}%
      \hfil}}%
\epsfxsize=0pt\epsfysize=0pt}%
%
%
{\catcode`\%=12 \global\let\epsfpercent=
%
%
\long\def\epsfaux#1#2:#3\\{\ifx#1\epsfpercent
   \def\testit{#2}\ifx\testit\epsfbblit
      \epsfgrab #3 . . . \\%
      \epsffileokfalse
      \global\epsfbbfoundtrue
   \fi\else\ifx#1\par\else\epsffileokfalse\fi\fi}%
%
%
\def\epsfgrab #1 #2 #3 #4 #5\\{%
   \global\def\epsfllx{#1}\ifx\epsfllx\empty
      \epsfgrab #2 #3 #4 #5 .\\\else
   \global\def\epsflly{#2}%
   \global\def\epsfurx{#3}\global\def\epsfury{#4}\fi}%
%
%
\def\epsfsize#1#2{\epsfxsize}%
%
%

\epsfverbosetrue			
\abovefigskip=\baselineskip		
\belowfigskip=\baselineskip		
\global\let\figtitlefont\bf		
\global\figtitleskip=.5\baselineskip	

\hbox{}
\hs-1.5 \vbox{\epsfig 1\hsize; 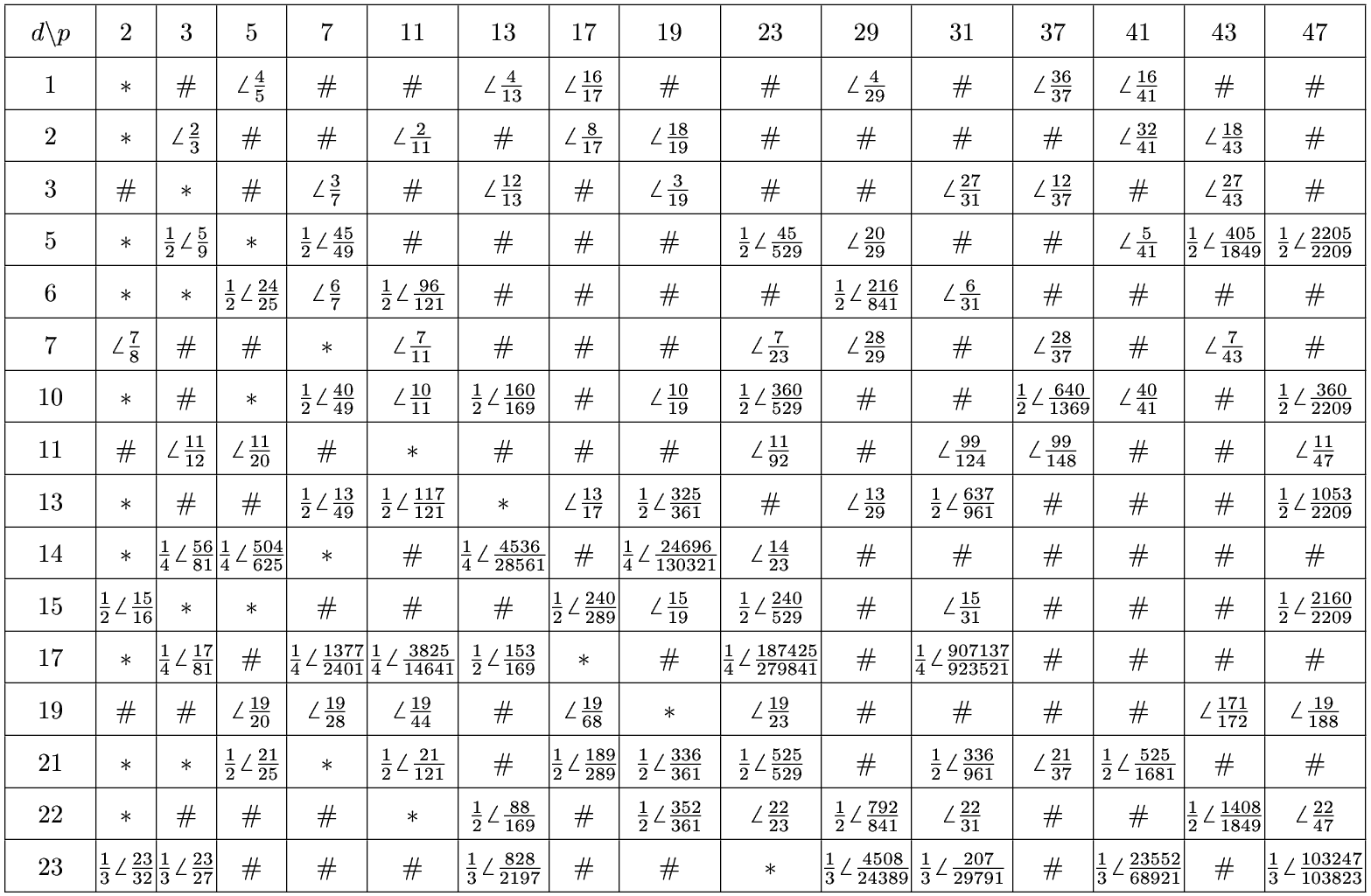;}
\vs-1.5
\nd Table 1.\ \ \ Basis elements $\langle p\rangle_d$ for some $p$ and $d$. 
$\#$ indicates that $(p)$ is prime in $\O_d$, while * indicates that $(p)$
ramifies.
\vfill \eject
\hbox{}
\vbox{\epsfig 1\hsize; 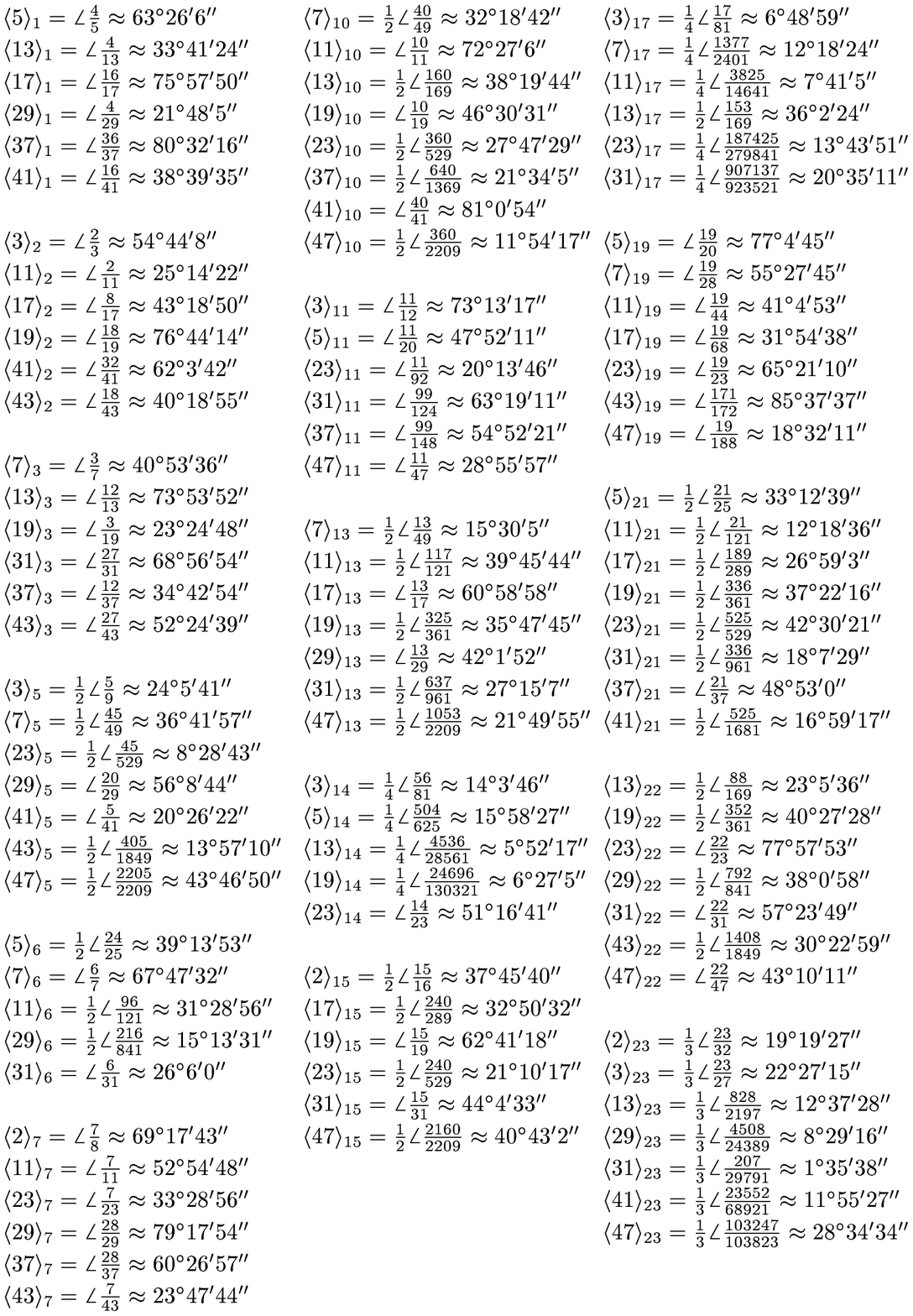;}
\vs0
\centerline{Table 2.\ \ \ The values of some basis elements $\langle p\rangle_d$.}
\vfill \eject
\nd \hbox{} \vs1
$$\vbox{\settabs 2 \columns
\+
Tetrahedron& $ - 12\langle 3\rangle_2$\cr
\+
Truncated tetrahedron& $  12\langle 3\rangle_2$\cr
\+
&\cr
\+
Cube& $0$\cr
\+
Truncated cube& $ - 24\langle 3\rangle_2$\cr
\+
&\cr
\+
Octahedron& $24\langle 3\rangle_2$\cr
\+
Truncated octahedron& $o$\cr
\+
Rhombicuboctahedron& $ - 24\langle 3\rangle_2$\cr
\+
&\cr
\+
Cuboctahedron& $- 24\langle 3\rangle_2$\cr
\+
Truncated cuboctahedron& $0$\cr
\+
&\cr
\+
Icosahedron& $  60\langle 3\rangle_5$\cr
\+
Truncated icosahedron& $  30\langle 5\rangle_1$\cr
\+
&\cr
\+
Dodecahedron& $-30\langle 5\rangle_1$\cr
\+
Truncated dodecahedron& $ - 60\langle 3\rangle_5$\cr
\+
Rhombicosidodecahedron& $  60\langle 3\rangle_5 - 30\langle 5\rangle_1$\cr
\+
&\cr
\+
Icosidodecahedron& $ - 60\langle 3\rangle_5 + 30\langle 5\rangle_1$\cr
\+
Truncated icosidodecahedron& $0$\cr
}$$
\vs1
\centerline{Table 3.\ \ \ The Dehn invariant for the non-snub unit edge Archimedean 
polyhedra.}
\vfill \eject
\input xy
\xyoption{poly}
\parskip=0 pt
\hbox{}\vs1
\hs1
\xy
/r10pc/ :{\xypolygon3{}}
\endxy
\vs-3.3 \hs2.35
${\tau\over 2}, {1\over 2}, {\sigma\over 2}, 0$
\vs3 \hs.3 ${\sqrt{5} \over 3}, {1 \over 3}$ \hs3.9 $ {\sqrt 5\over 5}$ 
\vs-2 \hs.8 ${\tau\over \sqrt{3}}, {1\over \sqrt{3}}, {\sigma\over \sqrt{3}}, 0$ 
\hs1.9 $\sqrt{{\tau\over \sqrt{5}}}, \sqrt{{\sigma\over \sqrt{5}}}, 0$
\vs1.9 \hs2.1 $\sqrt{{\tau^3\over 3\sqrt{5}}}, \sqrt{{\sigma^3\over 3\sqrt{5}}}$ 
\vs-3 \hs2.52 dyad
\vs1.8 \hs1.4 triad \hs1.8 pentad

\vs2 \nd
Figure 1.\ \ \ Cosines of angles between axes of fixed rotational
symmetry (shown at corners), and between axes of different rotational
symmetry (shown at edges).
\vfill \eject
\hbox{}\vs1
\hs1
\xy
/r10pc/
:{\xypolygon3{}} 
\endxy
\vs-3.3 \hs2.35
${\pi\over 5}, {\pi\over 3}, {2\pi\over 5}, {\pi\over 2}$
\vs2.7 \hs.1
${\pi\over 2}-2\langle 3\rangle_5,$
\vs.1 \hs.2
$\pi -2\langle 3\rangle_2$
\hs3.7 $ \langle 5\rangle_1$
\vs-2 \hs0 ${\pi\over 4}-\langle 3\rangle_5,\langle 3\rangle_2,
{\pi\over 4}+\langle 3\rangle_5, {\pi\over 2}$
\hs1.7 ${1\over 2}\langle 5\rangle_1, {\pi\over 2}-{1\over 2}\langle 5\rangle_1, 
{\pi\over 2}$
\vs1.9 \hs1.4 ${1\over 4}\pi+\langle 3\rangle_5-{1\over 2}\langle 5\rangle_1, 
{3\over 4}\pi-\langle 3\rangle_5-{1\over 2}\langle 5\rangle_1$
\vs-3 \hs2.52 dyad
\vs2 \hs1.4 triad \hs1.8 pentad

\vs2 \nd
Figure 2.\ \ \ Angles between axes of fixed rotational symmetry (shown
at corners), and between axes of different rotational symmetry (shown
at edges).
\vfill

\end